\documentclass[twocolumn,trackchanges]{aastex61}

\def\dotdeg{\hbox{$.\!\!^\circ$}}
\def\arcdeg{\hbox{$^\circ$}}

\def\arcsec{\hbox{$^{\prime\prime}$}}
\def\kms{\hbox{km s$^{-1}$}}
\def\VLSR{\hbox{$V_{\rm LSR}$}}
\def\Ekin{\hbox{$E_{\rm kin}$}}
\def\Vexp{\hbox{$V_{\rm exp}$}}
\def\texp{\hbox{$t_{\rm exp}$}}
\def\sun{\hbox{$_{\odot}$}}
\def\Leqplus{\hbox{$l\!=\!+1\fdg3$}}
\def\Leqminus{\hbox{$l\!=\!-1\fdg2$}}

\received{May 29, 2017}
\revised{}
\accepted{\today}

\submitjournal{ApJ}

\shorttitle{Another Bubble in the Central Molecular Zone of Our Galaxy}
\shortauthors{Tsujimoto et al.}

\begin{document}

\title{Detection of Another Molecular Bubble in the Galactic Center}


\author{Shiho Tsujimoto}
\affil{School of Fundamental Science and Technology, Graduate School of Science and Technology, Keio University, 3-14-1 Hiyoshi, Yokohama, Kanagawa 223-8522, Japan}

\author{Tomoharu Oka}
\affiliation{School of Fundamental Science and Technology, Graduate School of Science and Technology, Keio University, 3-14-1 Hiyoshi, Yokohama, Kanagawa 223-8522, Japan}
\affiliation{Department of Physics, Institute of Science and Technology, Keio University, 3-14-1 Hiyoshi, Yokohama, Kanagawa 223-8522, Japan}

\author{Shunya Takekawa}
\affiliation{School of Fundamental Science and Technology, Graduate School of Science and Technology, Keio University, 3-14-1 Hiyoshi, Yokohama, Kanagawa 223-8522, Japan}

\author{Masaya Yamada}
\affiliation{School of Fundamental Science and Technology, Graduate School of Science and Technology, Keio University, 3-14-1 Hiyoshi, Yokohama, Kanagawa 223-8522, Japan}

\author{Sekito Tokuyama}
\affiliation{School of Fundamental Science and Technology, Graduate School of Science and Technology, Keio University, 3-14-1 Hiyoshi, Yokohama, Kanagawa 223-8522, Japan}

\author{Yuhei Iwata}
\affiliation{School of Fundamental Science and Technology, Graduate School of Science and Technology, Keio University, 3-14-1 Hiyoshi, Yokohama, Kanagawa 223-8522, Japan}

\author{Justin A. Roll}
\affiliation{School of Fundamental Science and Technology, Graduate School of Science and Technology, Keio University, 3-14-1 Hiyoshi, Yokohama, Kanagawa 223-8522, Japan}

\begin{abstract}

The $\Leqminus$ region in the Galactic center has a high CO {\it J}=3--2/{\it J}=1--0 intensity ratio and extremely broad velocity width.  This paper reports the detection of five expanding shells in the $\Leqminus$ region based on the CO {\it J}=1--0, $^{13}$CO {\it J}=1--0, CO {\it J}=3--2, and SiO {\it J}=8--7 line data sets obtained with the Nobeyama Radio Observatory 45 m telescope and James Clerk Maxwell Telescope.  The kinetic energy and expansion time of the expanding shells are estimated to be $10^{48.3\mbox{--}50.8}$ erg and $10^{4.7\mbox{--}5.0}$ yr, respectively.  {The origin of these expanding shells is discussed.}  The total kinetic energy of $10^{51}$ erg and the typical expansion time of $\sim\!10^5$ yr {correspond to} multiple supernova explosions at a rate of $10^{-5}$--$10^{-4}$ yr$^{-1}$.  This indicates that the $\Leqminus$ region may be a molecular bubble associated with an embedded massive star cluster{, although the absence of an infrared counterpart makes this interpretation somewhat controversial.} The expansion time of the shells increases as the Galactic longitude decreases, suggesting that the massive star cluster is moving from Galactic west to east with respect to the interacting molecular gas.  {We propose a model wherein the} cluster is moving along the innermost $x_1$ orbit {and} the interacting gas collides with it from the {Galactic} eastern side.  
\end{abstract}
\keywords{Galaxy: center --- ISM: clouds --- ISM: molecules}

\section{Introduction} \label{sec:intro}
\begin{figure*}
\begin{center}
\includegraphics{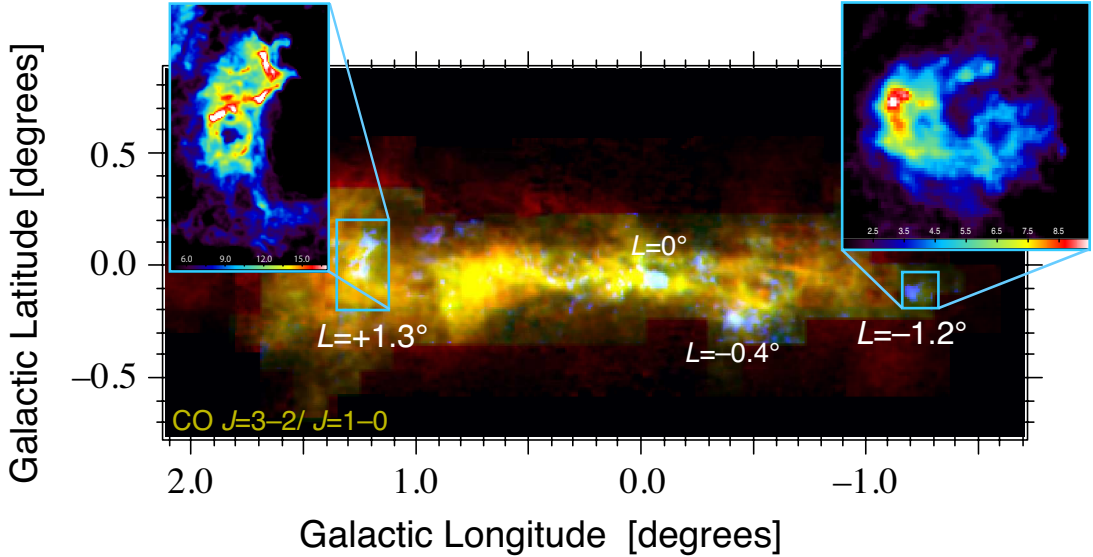}
\caption{Composite CO image of the Galactic CMZ.  Red and green indicate the velocity-integrated intensities of the CO {\it J}=1--0 and CO {\it J}=3--2 lines, respectively and blue is the CO {\it J}=3--2 emission integrated for data with high CO {\it J}=3--2/{\it J}=1--0 intensity ratios ($R_{3\mbox{--}2/1\mbox{--}0}\ge1.5$).  The two zoomed-up maps show the velocity-integrated CO {\it J}=3--2 line emission around the $\Leqplus$ and $\Leqminus$ regions.}
\label{fig:f1}
\end{center}
\end{figure*}

The central molecular zone (CMZ), the region within $\!\sim\!200$ pc from the center of our Galaxy, is characterized by a large amount of dense [$n\left({\rm H_2}\right)\!\ge\!10^4\;{\rm cm^{-3}}$] and warm ($T_{\rm k}\!=\!30\mbox{--}60\;{\rm K}$) molecular gas \citep{Morris83,Paglione98}, having highly turbulent and complex kinematics, with a large velocity width ($\Delta V\!\ge\!20$ \kms).  Our group surveyed the CMZ in the CO {\it J}=1--0 line (115 GHz) covering $-1\fdg5\le l\le+3\fdg4,\,|b|\le0\fdg6$ by using the Nobeyama Radio Observatory (NRO) 45 m telescope \citep{Oka98b} and the CO {\it J}=3--2 line (346 GHz) covering $-1\fdg5\!\le\!l\!\le\!+1\fdg0,\,|b|\!\le\!0\fdg2$ by using the Atacama Submillimeter Telescope Experiment (ASTE: \citealp{Oka07,Oka12}). By carefully inspecting these data sets, we found four high CO {\it J}=3--2/{\it J}=1--0 intensity ratio ($R_{3\mbox{--}2/1\mbox{--}0}\ge1.5$) regions, at the Galactic longitudes; $\!l=\!+1\fdg3,\,0\arcdeg,\,\mbox{--}0\fdg4,\,{\rm and}\,\mbox{--}1\fdg2$ (Figure \ref{fig:f1}).  All of the high-ratio regions show extraordinary broad velocity widths ($\Delta V\!\ge\!50$ \kms), containing several high-velocity compact clouds (HVCCs), which are a peculiar population of molecular clouds with compact appearances ($d<10\;{\rm pc}$) and broad velocity widths ($\Delta V_{\rm LSR}\ge 50\; {\rm km\,s^{-1}}$) identified in the CMZ \citep{Oka99, Oka01, Oka07, Tanaka07, Oka12}. 

The $\Leqplus$ region, one of the high $R_{3\mbox{--}2/1\mbox{--}0}$ regions, contains nine expanding shells, which typically have very high kinetic energy reaching $\Ekin\!\sim\!10^{53}$ erg, and an expansion time $\texp\!\sim\!10^{5}$ yr.  Most importantly, two of the expanding shells are associated with SiO clumps at their high-velocity ends.  As gas-phase SiO is a well-established probe of strong interstellar shocks, a series of supernova (SN) explosions may be responsible for the formation of the expanding shells.  The kinetic energy requires tens or hundreds of SN explosions.  Combined with the typical expansion time of the shells, $10^5$ yr, the SN rate should be $10^{-4}$--$10^{-3}$ yr$^{-1}$.  Thus we suggest that the $\Leqplus$ region is in an early stage of superbubble formation and that a very massive ($\sim\!10^{5\mbox{--}6}\,M\sun$) star cluster is embedded in it \citep{Tanaka07}.  

Another high $R_{3\mbox{--}2/1\mbox{--}0}$ region, that is, the $\Leqminus$ region, shares the principal properties with the $\Leqplus$ region.  The locations of these two high-ratio regions in the sky plane are roughly symmetric with respect to the Galactic center.  The $\Leqminus$ region contains an expanding shell which has a size of $\sim\!10$ pc, showing an extraordinary broad velocity width ($\Delta V\!\geq\!100$ \kms) \citep{Oka12}.  In this paper, we report the detection of four additional expanding shells in the $\Leqminus$ region based on the newly obtained high-resolution CO survey data. The $\Leqminus$ region is another candidate for a molecular bubble in the CMZ of our Galaxy. Throughout this paper, we adopt 8.3 kpc as the distance to the Galactic center.  

\section{Data} \label{sec:obs}
{The mapping data were obtained by using the NRO 45 m radio telescope and the James Clerk Maxwell Telescope (JCMT). The observed lines are $^{12}$CO {\it J}=1--0, $^{13}$CO {\it J}=1--0, $^{12}$CO {\it J}=3--2, and SiO {\it J}=8--7.
}

\subsection{NRO 45 m Data} \label{subsec:nro_obs}
The on-the-fly (OTF) mapping observation of $^{12}$CO {\it J}=1--0 (115.27120 GHz) { and} $^{13}$CO {\it J}=1--0 (110.20135 GHz) lines was performed using the NRO 45 m radio telescope in 2016 January--March and May {as parts of the large-scale surveys of the Galactic CMZ \citep{Tokuyama17}. To investigate the $\Leqminus$ region, we mainly analyzed a $9\farcm6\!\times\!9\farcm6$ area ($-1\fdg32\!\le\! l\!\le\!-1\fdg16$ and $-0\fdg20\!\le\! b\!\le\!-0\fdg04$).} We employed the FOREST receiver operated in the two-sideband mode.  At 110 GHz, the half-power beamwidth (HPBW) of the telescope was $14\arcsec$ and the main beam efficiency ($\eta_{\rm MB}$) was 0.4.  We used the SAM45 spectrometers in the 1000 MHz bandwidth (244.14 kHz resolution) mode.  The system noise temperature ($T_{\rm sys}$) ranged from 300 to 800 K during the $^{13}$CO observations, and from 1000 to 2000 K during the $^{12}$CO observations.  Antenna temperatures were calibrated using the standard chopper-wheel method to obtain the $T_{\rm A}^*$ scale.  We scaled $T_{\rm A}^*$ by multiplying with $1/\eta_{\rm MB}$ to obtain the main-beam temperature ($T_{\rm MB}$).  All spectra were obtained through position switching between target scans and the clean reference position $(l,\,b)\!=\!(+2\arcdeg,\,+1\arcdeg)$.  The pointing errors were corrected every 1.5 hr by observing the SiO maser source VX Sgr at 43 GHz.  The pointing accuracy was better than $3\arcsec$ in both azimuth and elevation.  All the NRO 45 m data were reduced using the {\it NOSTAR}\footnote{\url{http://www.nro.nao.ac.jp/~nro45mrt/html/obs/otf/export-e.html}} reduction package. We subtracted the baselines of all spectra through linear fitting.

\subsection{JCMT Data}
The $^{12}$CO {\it J}=3--2 line (345.79599 GHz) and the SiO {\it J}=8--7 line (347.33058 GHz) data were obtained by the JCMT. The $^{12}$CO {\it J}=3--2 line observations were performed on 2013 September by the JCMT Plane Survey (JPS) team \citep{Parsons17} as part of the Galactic plane survey. To investigate the $\Leqminus$ region, we mainly analyzed a $9\farcm6\!\times\!9\farcm6$ area ($-1\fdg32\!\le\! l\!\le\!-1\fdg16$ and $-0\fdg20\!\le\! b\!\le\!-0\fdg04$).

We also observed the SiO {\it J}=8--7 line on 2017 June. We set a $9\farcm6\!\times\!9\farcm6$ area centered at $(l,\,b)\!=\!(-1\fdg24,\,-0\fdg12)$ as the mapping area, and $(l,\,b)\!=\!(-1\fdg24,\,-0\fdg62)$ as the clean reference position. The pointing errors were corrected by observing the standard pointing sources of the JCMT.

We used a single-sideband receiver, the Heterodyne Array Receiver Program (HARP; \citealt{Buckle09}). At 345 GHz, the HPBW of the telescope is $14\arcsec$, and the $\eta_{\rm MB}$ is $0.64$.  The ACSIS spectrometer was operated in the 1000 MHz bandwidth (976.56 kHz resolution) mode.  The $T_{\rm sys}$ ranged from 100 to 200 K during the HARP observations. The JCMT data were reduced with the {\it Starlink}\footnote{\url{http://starlink.eao.hawaii.edu/starlink}}software package.

\section{Results} \label{sec:result}
\subsection{Spatial Distribution} \label{subsec:spatial}
\begin{figure*}
\begin{center}
\includegraphics[scale=0.25]{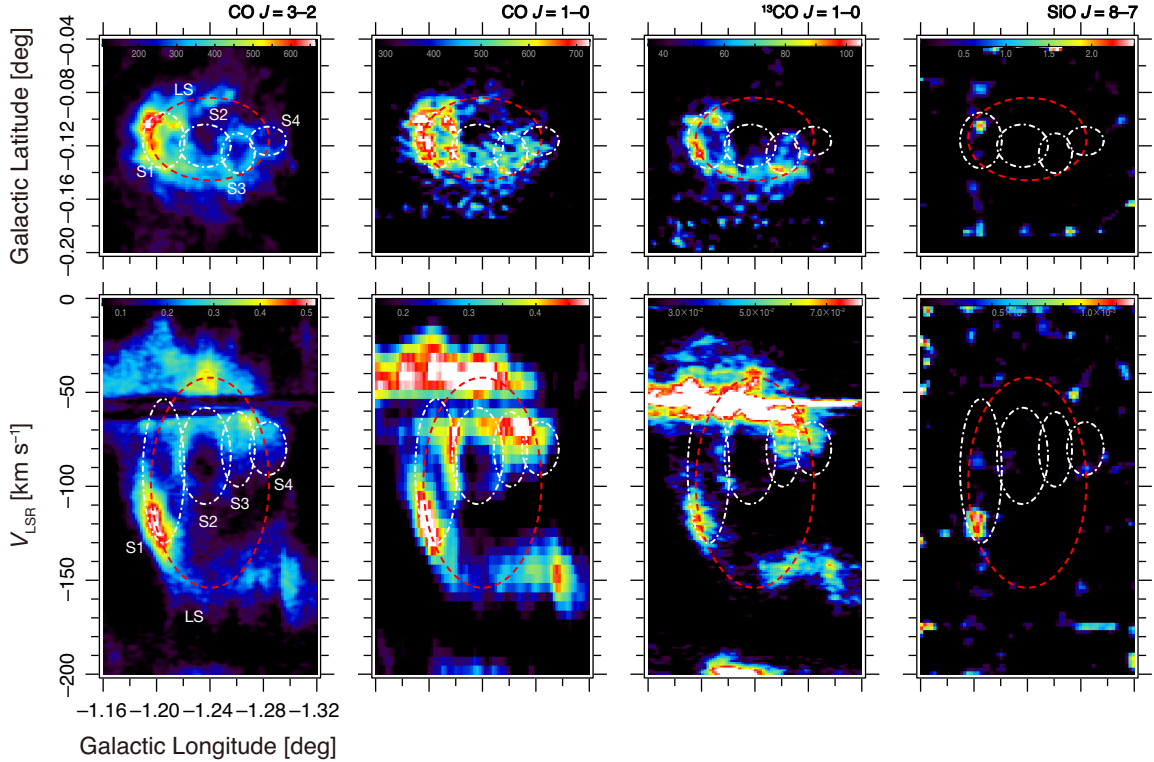}
\caption{The top four maps are the velocity-integrated ($\VLSR=-130$ to $-60\,\kms$) {\em l-b} maps and the bottom four are the {\em l-V} maps integrated over latitudes $b=-0\fdg14\,{\rm to}\,-0\fdg10$. From left to right, we see the CO {\it J}=3--2, CO {\it J}=1--0, $^{13}$CO {\it J}=1--0 and SiO {\it J}=8--7 maps.  The red-dashed ellipse shows the main large shell (LS), the white dashed-and-dotted ellipses show the small shells (S1--4).}
\label{fig:f2}
\end{center}
\end{figure*}

The top four panels of Figure \ref{fig:f2} show the velocity-integrated {\em l-b} maps of CO {\it J}=1--0, CO {\it J}=3--2, $^{13}$CO {\it J}=1--0, and SiO {\it J}=8--7 line emissions. The velocity range for integration is from $\VLSR\!=\!-130$ to $-60$ \kms.  First, all {\em l-b} maps of CO lines show an elliptical morphology with a less intense central cavity; we named this as the ``large shell (LS)''. The major axis of the LS is parallel to the Galactic plane. Velocity-integrated intensities are asymmetric, that is, they are very intense at the Galactic eastern side, and fainter at the western side.  These are several intense clumps, in which the peak intensity of CO {\it J}=3--2 line reaches almost 19 K at $(l,\:b)\!=\!(-1\fdg20,\,-0\fdg11)$, while the typical intensity of the shell is 5--10 K (Figure \ref{fig:f2}).  

In the velocity-integrated maps, especially in the CO {\it J}=3--2 map, we also noticed four small shells inside or at the edge of the LS.  We named them the ``small shells (S1--4)'', which are traced by the white dashed-and-dotted ellipses, while the LS is traced by the red-dashed ellipse (Figure \ref{fig:f2}). The most intense position of the CO lines is at a location where the LS and S1 overlap, and corresponds to the only position where the SiO {\it J}=8--7 line was detected. The centers of these shells are aligned roughly parallel to the Galactic plane, ranging over 11.6 pc in projected distance. The positions and sizes of these shells $(l,\,b,\,\Delta l,\,\Delta b)$ are summarized in Table \ref{tab:t1}.

\begin{table*}
\begin{center}
\caption{Properties of Expanding Shells}\label{tab:t1}
\begin{tabular}{cccccccccc}\hline\hline
Name&	{\em l}	&	{\em b}	&$\Delta l$&$\Delta b$&$V_{\rm exp}$&$M$&${\rm log}_{10}\,E_{\rm kin}$&$t_{\rm exp}$	&${\rm log}_{10}\,P_{\rm kin}$ \\
	&	(degree)		&	(degree)		&	(pc)	&	(pc)	&	(\kms)	&	$(10^3\,M\sun)$&	(erg)		&	($10^4$ year)&	(erg s$^{-1}$)			\\ \hline
LS	&	$-1.24$	&	$-0.12$	&	13	&	9.1	&	60		&	19.0			&	50.8		&	8.6		&			38.4			\\
S1	&	$-1.21$	&	$-0.12$	&	4.3	&	5.8	&	40		&	3.81			&	49.8		&	5.9		&			37.5			\\
S2	&	$-1.24$	&	$-0.12$	&	5.6	&	6.0	&	35		&	2.06			&	49.4		&	7.8		&			37.0			\\
S3	&	$-1.26$	&	$-0.13$	&	3.6	&	4.4	&	20		&	1.49			&	48.8		&	9.5		&			36.3			\\
S4	&	$-1.29$	&	$-0.12$	&	4.0	&	3.1	&	15		&	0.99			&	48.3		&	11.2		&			35.8			\\ \hline
Total	&			&			&		&		&			&	27.3			&	50.9		&			&			38.5			\\ \hline	
\end{tabular}
\end{center}
\end{table*}

\subsection{Kinematics} \label{subsec:kine}
The bottom four panels of Figure \ref{fig:f2} show the longitude--velocity ({\em l-V}) maps of the observed lines integrated over latitude.  The latitude range for the integration was from $b\!=\!-0\fdg14$ to $-0\fdg10$. The straight horizontal lines at $\VLSR\!\simeq\!-60\,\kms$ on the {\em l-V} maps are absorption features due to the spiral arm in the Galactic disk (3 kpc arm). The emission at the $\VLSR\!\simeq\!-200\,\kms$ represents the expanding molecular ring (EMR; \citealt{Kaifu72, Scoville72}).  In addition, another wavy emission feature appears at $\VLSR\!\simeq\!-150$ \kms. 

The LS and S1--4 can be clearly traced as parts of ellipses in these {\em l-V} maps, especially in the CO {\it J}=1--0 and {\it J}=3--2 maps. The elliptical shapes in the position--velocity maps indicate expanding motion. The expansion velocities ($\Vexp$) of the small shells are slower than those of the LS. The expansion velocities of the shells are listed in Table \ref{tab:t1}. Line intensities are enhanced at locations where the shells overlap. The LS/S1 overlapping area shows especially intense emission with clumpy morphology in the {\em l-b} space. This corresponds to the Galactic eastern edge of the LS and the S1, and the high-velocity ends of the S1.

\section{Discussion} \label{sec:discuss}

\subsection{Driving Mechanisms} \label{subsec:mechanism}
The {\em l-b} and {\em l-V} maps show incomplete elliptical shapes with broad velocity widths (Figure \ref{fig:f2}). {Several processes are proposed to explain the morphology and kinematics of the \Leqminus\ region; (1) cloud-cloud collision, (2) magnetically floated loop, and (3) multiple supernova explosions.}

\paragraph{Cloud-Cloud Collision}
A collision of clouds may generate a broad velocity width and elliptical spatial morphology \citep{Habe92}.  In this case, position-velocity maps should show less intense emission bridges which connect two intense components with different velocities \citep{Matsumura12, Haworth15, Torii17}.  {The $\Leqminus$ region has two velocity components at $\VLSR\!\sim\!-50\,\kms$ and $-150\,\kms$, and a broad velocity width feature at $l\sim\!-1\dotdeg 20$ seems to connect these components.  Cloud-cloud collisions may frequently occur at the interface between $x_1$ and $x_2$ orbits, as well as at the inner ends of the dust lanes where the `spray' of incoming gas collides with the innermost $x_1$ orbit \citep{Binney91}.  These collisions definitely generate broad velocity width features in position-velocity maps (e.g., \citealt{Liszt06}). }

\paragraph{Magnetically Floated Loop}
We may expect broad velocity width features at two footpoints of the molecular loops.   There are two molecular loops, loop 1 and loop 2, discovered in the Galactic center region \citep{Fukui06}.  {They are interpreted as formed by magnetic floatation caused by Parker instability.  Large velocity dispersion due to the collision of gas at each footpoint is expected by the magnetic loop model.}  We note that the western edge of LS is at $l\sim\!-1\fdg 3$ and the eastern footpoint of the loop 1 is at $l\sim\!-2\arcdeg$, {indicating no physical connection between them, and no molecular loop structure has been} found above/below the $\Leqminus$ region.

\paragraph{Multiple Supernova Explosions}
{The interaction with a supernova explosion naturally explains the elliptical morphology, broad velocity width, and expanding shell kinematics of the molecular gas.  The multiple shells may indicate multiple supernova explosions in the $\Leqminus$ region within their expansion times.  Although there remains some subjectivity in the shell identification, this interpretation can explain the observational facts most naturally.  The detection of the SiO {\it J}=8--7 line, which is a good tracer of shocked molecular gas, at the high-velocity end of the S1 may support the origin related to a local explosive event.  Hereafter we discuss the $\Leqminus$ region basically along with the multiple supernova explosion model.}

\subsection{Physical Parameters} \label{subsec:phys_para}
Here, we estimate the physical parameters of the shells.  In principle, the gas mass ($M$) can be derived  by summing up the column density of gas in each shell.  Some shells overlap in the line-of-site direction, and it is difficult to separate individual shells in the data cube.  Thus, we simply define each shell as a finite thickness ellipsoid in the {\it l-b-V} space.  Pixels in an overlapping area belong to shells redundantly.  We suppose that all gas in the $\Leqminus$ region belongs to the five shells. The gas column density is calculated from the CO {\it J}=1--0 data by assuming a local thermodynamic equilibrium (LTE). The optical depth ($\tau=2.5$) and the excitation temperature ($T_{\rm ex}=10$ K) are determined based on the CO {\it J}=1--0 and $^{13}$CO {\it J}=1--0 data. We adopt $[{\rm CO}]/[^{13}{\rm CO}]\!=\!24$, $[{\rm H_2}]/[{\rm CO}]\!=\!10^4$ \citep{Langer90}.

The kinetic energy $(E_{\rm kin})$ of each shell is calculated by 
\begin{eqnarray}
\Ekin=\frac{1}{2}MV_{\rm exp}^2\,.
\end{eqnarray}
\noindent The expansion time of a shell is derived according to its size and expansion velocity.  By considering the asymmetric shape of the shells, we define the size of a shell as
\begin{eqnarray}
R=\sqrt{\frac{\Delta l}{2}\frac{\Delta b}{2}}\,,
\end{eqnarray}
\noindent and thus $t_{\rm exp}$ is calculated as 
\begin{eqnarray}
t_{\rm exp}=\frac{R}{V_{\rm exp}}\,.
\end{eqnarray}
The calculated physical parameters, $M$, $\Ekin$, $t_{\rm exp}$, and kinetic power ($P_{\rm kin}=\Ekin/t_{\rm exp}$) are listed in Table \ref{tab:t1}. Such a large kinetic power implies the close association of an energy source to this cloud.

\subsection{Energy Source of the $\Leqminus$ Region}
The total kinetic energy of the five expanding shells amounts to $10^{50.9}$ erg. By considering the typical baryonic energy supplied by an SN explosion, $E_{\rm bar}\!\sim\!10^{51}$ erg \citep{Chevalier74}, the energy conversion efficiency ($\eta\!=\!\Ekin/E_{\rm bar}$) to the gas kinetic energy is as low as $\sim\!0.1\mbox{--}0.3$ \citep{Chevalier74,Salpeter76}, and the total kinetic energy corresponds to $\sim\!10^{0.4\mbox{--}0.9}$ SNe.  Combined with the typical expansion time of the shells ($\sim\!10^5$ yr), the SN rate is estimated to be $\sim\!10^{-5}\mbox{--}10^{-4}$ yr$^{-1}$.  This high SN rate within the small volume of the $\Leqminus$ region indicates that a  young massive star cluster is embedded here. The absence of H$_{\rm II}$ regions indicates that the cluster age is older than 10 Myr, the main sequence lifetime of a $15\,M\sun$ star. The presence of a type-II SN indicates a cluster age of less than 30 Myr.   

If we assume a uniform stellar age (instantaneous star formation), Scalo initial mass function (IMF; \citealt{Scalo86}) with lower/higher mass cutoffs of $0.08\,M\sun/100\,M\sun$, and mass of heaviest stars of 8--15 $M\sun$, the cluster mass is estimated to be $M_{\rm cl}\!\sim\!10^{5.6}\,\eta^{-1}\,M\sun$. This is two orders of magnitude larger than those of the Arches and Quintuplet clusters, $\sim10^4\,M\sun$ \citep{Figer99}, and is an order of magnitude lower than that of the cluster embedded in the $\Leqplus$ region (Table \ref{tab:t2}). 

The absence of an infrared counterpart seems to be at variance with the cluster-origin model.  In fact, the total luminosity of the young $M_{\rm cl}\!\sim\!10^{5.6}\,\eta^{-1}\,M\sun$ cluster amounts to $L_{\rm cl}\!\sim\!10^{7.1}\,\eta^{-1}\,L\sun$ \citep{Williams94}, while the far-infrared luminosity of the $\Leqminus$ region is $\sim\!10^{5.9}\,L\sun$.  The order of magnitude discrepancy may indicate that the IMF in the hypothetical cluster is abnormal.  For instance, a shallower slope, as well as a higher low-mass cutoff of the IMF, significantly decrease the estimated cluster mass.  {Alternatively, an abundance of dark stellar remnants in the cluster could suggest other kinds of explosions, such as Type Ia SNe and neutron star mergers \citep{Rosswog13}. }

\begin{table}
\begin{center}
\caption{Properties of Gases and Embedded Star Cluster}\label{tab:t2}
\begin{tabular}{ccccc}\hline\hline
Region	&${\rm log}_{10}\,E_{\rm kin}$	&$t_{\rm exp}$	&${\rm log}_{10}\,P_{\rm kin}$	&	$M_{\rm cl}$	\\
		&		(erg)				&($10^5$ year)	&	(erg s$^{-1}$)			&	($M_{\odot}$)	\\ \hline
\Leqplus	&		52.6--53.0			&	4.5--5.3	&		39.5				&$10^{6.6}\,\eta^{-1}$\\
\Leqminus	&		50.9				&	0.6--1.1	&		38.5				&$10^{5.9}\,\eta^{-1}$\\ \hline	
\end{tabular}
\end{center}
\end{table}

\subsection{Kinematics of Gas and Cluster}
\begin{figure}
\begin{center}
\includegraphics[scale=0.47]{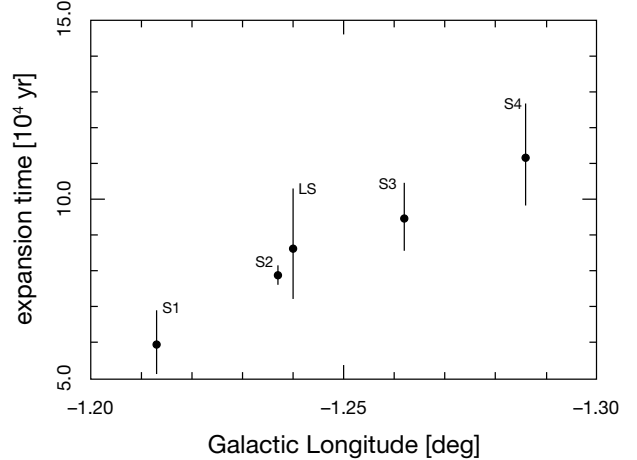}
\caption{Calculated expansion times plotted against the Galactic longitude of the center of each shell. The error bars were calculated by substituting $R$ with the major/minor radius of each shell.}
\label{fig:f3}
\end{center}
\end{figure}

Figure \ref{fig:f3} shows a plot of $t_{\rm exp}$ of the shells against the Galactic longitude ($l$).  The apparent gradient in $t_{\rm exp}$ along $l$ indicates a relative velocity of $\sim\!200$ \kms\ between the cluster and interacting gas.  This relative velocity is similar to that of the Galactic rotation.  The shorter $t_{\rm exp}$ in the larger longitude implies that the cluster is relatively moving toward a positive Galactic longitude relative to the interacting gas.  

\begin{figure}
\begin{center}
\includegraphics[scale=0.3]{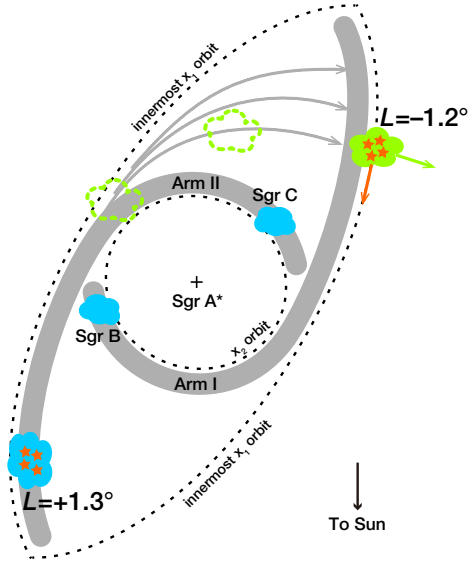}
\caption{Schematic face-on view of the CMZ. The black dashed line represents the innermost $x_{1}$ orbit, and the black cross shows the location of Sgr A*. The bottom left cyan object is the $\Leqplus$ complex, and the two cyan objects on arms I and II are the Sgr B and Sgr C regions, respectively. The orange arrow shows the direction along which the cluster is moving. The gray arrows show the direction along which the ``sprayed'' molecular gas is moving. The sprayed gas, shown as dotted and filled green objects, reaches to the other side of the innermost $x_{1}$ orbit and encounters a cluster.}
\label{fig:f4}
\end{center}
\end{figure}

This situation can be understood by considering the cluster and interacting gas on different orbits.  First, we assume Sofue's configuration of dense molecular gas \citep{Sofue95} and Binney's concept of gas kinematics in the CMZ \citep{Binney91}.  Large-scale CO {\it J}=1--0 data \citep{Oka98b} show that the $\Leqminus$ and $\Leqplus$ regions correspond to the extensions of Sofue's arms I and II, respectively.  Figure \ref{fig:f4} shows a schematic face-on view of the CMZ.  Accorging to Binney's concept, ``sprayed gas'' arises from locations where the innermost $x_1$ orbit and the outermost $x_2$ orbit intersect.  The sprayed gas flows into the inner orbits, and then collides with the other side of the innermost $x_1$ orbit to generate a pair of shock fronts.  If the hypothetical cluster exists on the innermost $x_1$ orbit, the presence of a high relative velocity between the cluster and interacting gas can be understood.  A parcel of sprayed gas might encounter the massive cluster, then receive expanding motions by SNe.  

The discovery of expanding shells at the longitude of $\sim\!-1\fdg2$ adds another molecular bubble in our Galaxy's CMZ.  The age of the embedded cluster is similar to that of the cluster in the $\Leqplus$ region, which may have been formed by the same mechanism as the $\Leqminus$ region.  This suggests that active star formation may have occurred at multiple places in the innermost $x_1$ orbit $\sim\!10^{7}$ yr ago.

\acknowledgments
We are grateful to the Nobeyama Radio Observatory (NRO) and James Clerk Maxwell Telescope (JCMT) staff for their excellent support of the observations. The NRO is a branch of the National Astronomical Observatory of Japan, National Institutes of Natural Sciences. The JCMT is operated by the East Asian Observatory on behalf of The National Astronomical Observatory of Japan, Academia Sinica Institute of Astronomy and Astrophysics, the Korea Astronomy and Space Science Institute, the National Astronomical Observatories of China, and the Chinese Academy of Sciences (Grant No. XDB09000000), with additional funding support from the Science and Technology Facilities Council of the United Kingdom and participating universities in the United Kingdom and Canada. { This research was supported by a grant from the Hayakawa Satio Fund awarded by the Astronomical Society of Japan.} T.O. acknowledges support from JSPS Grant-in-Aid for Scientific Research (B) No. 15H03643. S.T. acknowledges support from JSPS Grant-in-Aid for Research Fellow No. 15J04405.

\software{NOSTAR, Starlink\citep{Currie14}}


\begin{thebibliography}{99}
\bibitem[Binney et al.(1991)]{Binney91}Binney, J., Gerhard, O., E., Stark, A., A., Bally, J., and Uchida, K., I., 1991, MNRAS, 252, 210
\bibitem[Buckle et al.(2009)]{Buckle09}Buckle, J., V., Hills, R., E., Smith, H., et al., 2009, MNRAS 399, 1026
\bibitem[Chevalier(1974)]{Chevalier74}Chevalier, R., A., 1974, \apj, 188, 501
\bibitem[Currie et al.(2014)]{Currie14}Currie, M., J., Berry, D., S., Jennes, T., et al. 2014, ASPC, 485, 391
\bibitem[Figer et al.(1999)]{Figer99}Figer, D., F., Kim, S., S., Morris, M., et al., 1999, \apj, 525, 750
\bibitem[Fukui et al.(2006)]{Fukui06}Fukui, Y., Yamamoto, H., Fujishita, M., et al. 2006, Sci, 314, 106
\bibitem[Habe \& Ohta(1992)]{Habe92}Habe, A., and Ohta, K., 1992, PASJ, 44, 203
\bibitem[Haworth et al.(2015)]{Haworth15}Haworth, T., J., Tasker, E., J., Fukui, Y., et al., 2015, MNRAS, 450, 10
\bibitem[Kaifu et al.(1972)]{Kaifu72}Kaifu, N., Kato, T., and Iguchi, T., 1972, NPhS, 238, 105
\bibitem[Langer \& Penzias(1990)]{Langer90}Langer, W., D., and Penzias, A., A., 1990, \apj, 357, 477
{\bibitem[Liszt(2006)]{Liszt06}Liszt, H., S., 2006, A\&A, 447, 533}
\bibitem[Matsumura et al.(2012)]{Matsumura12}Matsumura, S., Oka, T., Tanaka, K., et al., 2012, \apj, 756, 87
\bibitem[Morris et al.(1983)]{Morris83}Morris, M., Polish, N., Zuckerman, B., and Kaifu, N., 1983, \aj, 88, 1228
\bibitem[Oka et al.(1998)]{Oka98b}Oka, T., Hasegawa, T., Sato, F., Tsuboi, M., and Miyazaki, A., 1998, \apjs, 118, 455
\bibitem[Oka et al.(1999)]{Oka99}Oka, T., White, G., J., Hasegawa, T., et al., 1999, ApJ, 515, 249
\bibitem[Oka et al.(2001)]{Oka01}Oka, T., Hasegawa, T., Sato, F., Tsuboi, M., and Miyazaki, A., 2001, PASJ, 53, 787
\bibitem[Oka et al.(2007)]{Oka07}Oka, T., Nagai, M., Kamegai, K., Tanaka, K., and Kuboi, N., 2007 PASJ, 59, 15
\bibitem[Oka et al.(2012)]{Oka12}Oka, T., Onodera, Y., Nagai, M., et al., 2012. \apjs, 201, 14
\bibitem[Paglione et al.(1998)]{Paglione98}Paglione, T., A., D., Jackson, J., M., Bolatto, A., D., and Heyer, M., H., 1998, \apj, 493, 680
\bibitem[Parsons et al.(2017)]{Parsons17}Parsons, H., Dempsey, J., T., Thomas, H., S., et al. 2017, {\apjs, 234, 22}
{\bibitem[Rosswog et al.(2013)]{Rosswog13}Rosswog, S., Piran, T., and Nakar, E., 2013, MNRAS, 430, 2585}
\bibitem[Salpeter(1976)]{Salpeter76}Salpeter, E., E., 1976, \apj, 206, 673
\bibitem[Scalo(1986)]{Scalo86}Scalo J.M., 1986, FCPh, 11, 1
\bibitem[Scoville(1972)]{Scoville72}Scoville, N., Z., 1972, \apjl, 175, L127
\bibitem[Sofue(1995)]{Sofue95}Sofue, Y., 1995, PASJ, 47, 527
\bibitem[Strickland \& Heckman(2009)]{Strickland09}Strickland, D. K., and Heckman, T. M. 2009, \apj, 697, 2030
\bibitem[Tanaka et al.(2007)]{Tanaka07}Tanaka, K., Kamegai, K., Nagai, M., and Oka, T., 2007, PASJ, 59, 323
\bibitem[Tokuyama et al.(in preparation)]{Tokuyama17}Tokuyama, S., Oka, T., Takekawa, S., et al., in prep.
\bibitem[Torii et al.(2017)]{Torii17}Torii, K., Hattori, Y., Hasegawa, K., et al., 2017, \apj, 835, 142
\bibitem[Williams \& Perry(1994)]{Williams94}Williams, R., J., R., Perry, J., J., 1994, MNRAS, 269, 538
\end{thebibliography}
\end{document}